# Direct in-situ observation of structural transition driven actuation in VO$_2$ utilizing electron transparent cantilevers

*B.Viswanath\* and Shriram Ramanathan*

Harvard School of Engineering and Applied Sciences, Cambridge, MA 02138, United States

Email: kbviswa@seas.harvard.edu  and shriram@seas.harvard.edu

Direct imaging and quantification of actuation in nanostructures that undergo structural phase transitions could advance our understanding of collective phenomena in the solid state. Here, we demonstrate visualization of structural phase transition induced actuation in a model correlated insulator vanadium dioxide by *in-situ* Fresnel contrast imaging of electron transparent cantilevers. We quantify abrupt, reversible cantilever motion occurring due to the stress relaxation across the structural transition from monoclinic to tetragonal phase with increasing temperature. Deflections measured in such nanoscale cantilevers can be directly correlated with macroscopic stress measurements by wafer curvature studies as well as temperature dependent electrical conduction allowing one to interrogate lattice dynamics across length scales.

## Introduction

Cantilevers with high sensitivity, high response speed and low noise are of growing interest in areas of nanoscale measurements and sensor devices. Electrical and optical methods are often utilized to investigate cantilever actuation with high precision. While the size reduction in the cantilever dimension is attractive for energy efficient nanoscale devices, progress in measuring actuation in smaller cantilevers is challenging as the resolution of optical methods are constrained by diffraction effects[1] and focusing.[2] Electrical methods based on piezoresistance of semiconducting materials could have restrictions due to reduced carrier density in low dimensional materials.[3] Developments in transmission electron microscopy (TEM) has enabled direct imaging of mechanical motion in nano and microscale structures[4,5], structural phase transitions[6], and interfacial phenomena with atomic scale resolution.[7-9] In this study, we present direct observation of actuation[10-12] in $VO_2$ during the thermally-driven phase transition by Fresnel contrast imaging in a transmission electron microscope utilizing electron transparent cantilever structures. $VO_2$ is a model correlated electron system that shows a sharp insulator-metal transition accompanying a structural transition. The material is of exceptional interest in condensed matter materials sciences towards understanding elementary mechanisms of the phase transition and exploring new devices enabled by reversible band gap closure that is a rare property. The structural phase transition in $VO_2$ involves monoclinic ($P2_1/c$) to tetragonal ($P4_2/mnm$) symmetry change at 68°C during heating.[13,14] The presence of V-V cation pairs along $a_m = 2c_r$ axis with alternate spacing of 0.265 nm and 0.312 nm is identified as a striking feature of monoclinic phase against the regular 0.287 nm spacing in the tetragonal phase.[15] The structural transition accompanying the electronic transition creates yet another opportunity to combine new functionality like in actuators, chemically driven transitions and environmentally aware materials in a broader setting. Recently, transformation stress in $VO_2$ is being explored for developing electromechanical actuators[10,11], cantilevers[16] and strain sensors.[17] In situ wafer curvature measurements have revealed large reversible transformation stress in crystalline $VO_2$ thin films across the phase transition over numerous thermal cycles.[12] Such transformation induced bending curvatures are also seen in $VO_2$ cantilevers on heating. New techniques to visualize and quantify nanoscale actuation would be of interest to further advancing our understanding of such structural transitions. Here we demonstrate the visualization of nanoscale cantilever motion, and quantification in the TEM and correlate directly to local structural transition as well as macroscopic wafer scale in-situ curvature measurements. This technique facilitates concurrent visualization and measurement of cantilever deflection along with structural investigations in an important model correlated oxide system where lattice distortions have profound influence on electronic properties. The approach presented here is of broad relevance to investigate structural dynamics and its actuation response in the solid state including in low-dimensional materials. The demonstrated visualization and quantification of actuation in small volume systems by TEM could have broad impact in studies of several classes of functional oxides, piezoelectric materials, shape memory alloys, bilayer structures and carbon nanotube based cantilevers that display reversible deflection.

## Results and Discussion

**Cantilever Fabrication and Imaging**

Fresnel contrast imaging has been shown to be an elegant technique to study interfaces and grain boundaries by through focal series of images, see for example the classic works of Clarke[18], Ross and Stobbs[19] and Stobbs et al.[20] Here, we apply Fresnel contrast technique for visualizing and measuring actuation by imaging electron transparent $VO_2$ cantilever edges in-situ spanning the phase transition boundary. A typical electron optics ray diagram is shown in Fig. 1a along with the experimental protocol to rigorously measure the cantilever deflection. The maximum cantilever tip deflection can be directly observed at the cantilever edge by tracing the defocus or specimen height as the temperature is varied. The presence of Fresnel fringes around the edge of cantilever enables measurement of deflection with few tens of nanometer precision. The interference effect between the transmitted wave passing through vacuum and diffracted wave from the edge of the cantilever with dissimilar potential varies based on the distance between the image plane and the objective lens focal plane. For a static specimen, no detectable fringes appear when the objective lens of the microscope is in exact focus with image plane. Varying the objective lens current moves the focus below (under focus) or above (over focus) the image plane leading to the appearance of bright or dark fringes respectively. Similarly, varying the specimen height below or above the eucentric position results in bright (under focus) or dark fringes (over focus) respectively for constant defocus. When the static specimen is replaced with a $VO_2$ cantilever, the temperature induced dynamic motion of the cantilever introduces continuous changes in Fresnel contrast across the phase transition enabling the visualization of actuation as the cantilever moves up or down. The actual cantilever deflection can be precisely monitored either by tracing the defocus value ($\Delta f$) or specimen height (z). For example, when the sample is heated from $61^oC$ to $74^oC$ (across the monoclinic-tetragonal phase boundary), the cantilever moves up causing an out of focus image at the image plane. It is possible to bring back the cantilever in to focus by exciting the objective lens (defocus) without physically changing the specimen height. On the other hand, one can also trace the deflection by varying the 'z' shift keeping the defocus ($\Delta f$) constant as the pole piece gap (~2 mm) provides enough room for specimen height adjustment. However the defocus method is more sensitive to Fresnel contrast and provides deflection measurement with resolution of sub-70 nm. SEM and TEM bright field images of $VO_2$ cantilevers are shown in Fig.1b and Fig.1c respectively. Inset in Fig. 1b show a secondary electron SEM image of cantilever recorded at high magnification showing the microstructure of $VO_2$ film. Cantilever dimensions were obtained from low magnification TEM images whereas the actual thicknesses of $VO_2$ layer and Si are determined from backscattered electron images from SEM. VO$_2$ cantilever of 5 μm x 2 μm dimensions with total thickness of 343 nm (t$_{Si}$ ~ 115 nm and t$_{VO2}$ ~ 228 nm) was used for in-situ TEM studies to probe the phase transition and visualize the cantilever deflection.

**Electrical and structural characteristics across the insulator-metal transition in VO$_2$**

Fig.2 shows electrical, structural and stress relaxation characteristics of insulator-metal transition in VO$_2$ thin films deposited on oxidized silicon wafers. While the resistance measurement provides information regarding the electrical component, the stress estimated from in-situ wafer curvature measurement and in-situ TEM studies allow probing of the structural component of the insulator-metal transition. Fig.2a shows the abrupt resistance and stress variation across the phase transition expected for a good quality VO$_2$ film on Si. The drastic reduction in resistance and abrupt tensile stress development[12] arise as a result of concurrent insulator to metal[21] and structural transitions.[22,23] The actual changes in the resistance and stress against temperature can be directly correlated to the fractions of insulating and metallic VO$_2$ phases from effective medium theory.[24] In order to probe the structural counterpart of the phase transition, TEM investigations were carried out from the same specimen. TEM bright field image recorded from the electron transparent polycrystalline VO$_2$/Si cantilever region is shown in Fig.2b. The average VO$_2$ grain size is on the order of ~100 nm and the typical ring pattern of monoclinic VO$_2$ is shown as inset. In situ bright field imaging across the transition shows abrupt changes in the contrast as the diffraction conditions are changing across the transition for constant tilting angle and specimen drift. While the bright field contrast is varying abruptly across the transition, there are no detectable changes such as appearance of internal boundaries. Typically, formation of new boundaries may be seen during reconstructive type phase transitions during heating.[25]

In order to capture the structural phase transition unambiguously, selected area electron diffraction pattern was recorded from single grains using the smallest aperture and shown in Fig.2c and Fig.2d at 21$^{o}$C and 115$^{o}$C respectively. The spot patterns are indexed with the monoclinic [-112] and tetragonal [-110] structures of same $VO_2$ grain below and above the transition temperature. The results are consistent with the simulated electron diffraction pattern using $VO_2$ monoclinic and tetragonal crystal structure parameters. Lattice images recorded at 22$^{o}$C and 110$^{o}$C (Fig.2e and Fig.2f) and the corresponding FFT patterns are shown as inset for reference. More importantly, abrupt variation (bright to dark) in the Fresnel contrast can be noticed at grain boundaries across the transition indicating the changes in the deflection of specimen as shown in Fig.3. In addition, step-like contrast variation is observed within the grain across the transition with appearance of thickness fringes at edges (Fig.3).

For the purpose of quantification of actuation and stress analysis, we use the cantilever edge for imaging at various temperatures using *in-situ* TEM Fresnel contrast. We also investigated the phase transition kinetics that is often performed in studies pertaining to understanding Martensitic-type transitions. Typically, the kinetics of martensitic transition is considered to be athermal in nature owing to the diffusionless process. Due to the large activation barrier, thermal fluctuations do not play a significant role in the athermal transition and are drive rate independent. Martensitic transitions in many systems show both athermal and isothermal components with respect to factors such as grain size, imperfections and heating rate, etc. For example, martensitic transition in $ZrO_2$ shows drive rate independent, athermal nature for coarse grained microstructure while the rate dependent, isothermal component of martensitic transition is observed for fine grained $ZrO_2$ below 100 nm.[26] The theory suggests that the nuceli cannot grow spontaneously to their full size when the crystallite size is less than the domian size of the new phase due to the strain energy arising due to the overlap of two phases. Hence the growth of second phase is inhibited by the grain boundaries in fine grained specimens. We note that the grain size of $VO_2$ thin film investigated here is ~ 100 nm and shows strong athermal characteristics. Any time dependence on the relative phase fractions was found to be negligible from the analysis and hence the changes are correlated with temperature change alone. Detailed *wafer curvature in-situ* stress measurements and resistance data probing the athermal kinetics of $VO_2$ thin films are shown in Fig.4**.** The overlapping plots of normalized resistance with initial time (time independent resistance) at different temperatures confirm the athermal nature of phase transition (Fig.4a). Similar time independent phase transition trend is observed in thin film stress measurements (Fig.4b). The effect of heating rate on the phase transition induced stress changes has been studied by varying the heating rate from $2^oC/min$ to $30^oC/min$. The transformation stress ($\Delta\sigma$ ~ 500 MPa) is nearly constant indicating complete phase transition in all the cases with different heating rates and shown in Fig.5 This clearly captures the heating rate independent stress relaxation of $VO_2$ phase transition. Typically, the rate dependence in martensitic transition is correlated to the annihilation rate of defects as imperfections obstruct the movement of dislocations and affect the kinetics of martensitic transition.[26] Rate independent athermal kinetics is observed primarily when the number of obstacles is very small that may be the case for high quality $VO_2$ films with good transition properties.

**Visualizing cantilever deflection through *in-situ* Fresnel contrast imaging**

The structural phase transition induces a large reversible tensile stress in $VO_2$ thin film of the order of several hundred MPa and the corresponding actuation in cantilever geometry can be directly observed by Fresnel contrast imaging. Fig.6 shows a series of bright field images from the $VO_2$ cantilever tip recorded at different temperatures. Initially the cantilever tip is brought in to focus and the sample is heated. At $25^{o}C$ there is negligible Fresnel contrast at the interface while the successive images recorded at different temperatures show development of a bright Fresnel fringe up to $61^{o}C$ followed by abrupt dark fringe development beginning from $64^{o}C$ (Fig.6). The systematic variation of Fresnel contrast is more evident from the line profile analysis of the corresponding images at different temperatures (Fig.6a). Line scan profiles are obtained perpendicular to the cantilever edge as marked XY in Fig.6b. Bright Fresnel fringe starts to appear at $35^{o}C$ and progressively increases as the temperature is raised up to $61^{o}C$. At $63^{o}C$, the peak intensity begins to decrease and the fringe contrast reverses from bright to dark proximal to structural phase transition. Across the phase transition boundary, the images are out of focus due to maximum cantilever tip displacement as seen from images as well as diffused line profiles above $65^{o}C$. In order to overcome the issue of out of focus imaging and to make precise deflection measurements across the phase transition, the image is iteratively brought into focus for every $2^{o}C$ temperature increment. As seen from the line plot, the bright fringe development during heating from room temperature to $61^{o}C$ indicates the downward deflection of cantilever. Strongly exciting the objective lens current (over focus) will account for such downward cantilever deflection. Whereas the observed dark fringe development across the phase transition ($62^{o}C$ to $74^{o}C$) corresponding to upward cantilever deflection require weak objective lens current (under focus) to bring the upward deflected cantilever in to focus. Fig.7 shows a series of representative TEM Fresnel contrast images of the $VO_2$ cantilever at different temperatures across the phase transition boundary before and after correcting the defocus values. For each and every temperature interval (2$^o$C), the cantilever goes out of focus whose magnitude was then precisely determined by varying the defocus with aid of Fresnel fringe seen at the cantilever edge. The same approach was used to monitor the deflection of cantilever against temperature from 25$^o$C to 100$^o$C during heating-cooling cycles. Representative defocus-temperature plot showing the abrupt defocus changes across the phase transition indicate the reversible cantilever oscillations with transformation induced defocus change of ~ 4 micron (Fig.8a). The magnitude and the nature of deflection were also confirmed over several heating-cooling cycles. Moreover, similar iterative experiment performed with "z" variation for the constant defocus also shows consistent cantilever deflection across the phase transition as plotted in Fig.8b. The normalized specimen height (stage) decreases across the transition as the cantilever moves up during heating (up to 61$^o$C). Note that variation in "z "or defocus is measured as a correction factor to reach focus from out of focused cantilever tip. Hence the specimen height (z) abruptly decreases to account for the out of focus caused by upward deflection of $VO_2$ cantilever during phase transition. On the other hand the "z" is increasing below and above the transition consistent with the trend seen in defocus variation across the phase transition.

The contribution from the overall specimen deflection has to be separated from just the cantilever deflection. In order to quantify the actual cantilever deflection, imaging and measurements were undertaken from the $VO_2$/Si cantilever base tracing the defocus variation and then the difference with respect to cantilever tip is plotted against temperature. The deflection of $VO_2$ cantilever during heating is shown in Fig.8c. Clearly, there are at least two different trends seen in the cantilever deflection at different stages. First, the cantilever tip shows maximum tip displacement (upward deflection) of 550 nm across the phase transition during heating. This is shown schematically as inset with upward cantilever deflection across the insulator-metal transition with different colors indicating the concurrent changes in resistance as shown in Fig.8c. The observed deflection of VO$_2$ cantilever is consistent with the trend seen in micro-scale VO$_2$ cantilevers of several hundred micron length measured by other methods.[16,27-29] *In-situ* Fresnel contrast imaging experiments were carried out on VO$_2$ films grown on aluminum (as another test specimen) also showed dark fringe development and deflection during heating across the insulator-metal transition. Second, the cantilever shows downward deflection trend both below and above the phase transition as evident from the bright fringe development in Fresnel contrast imaging. Note that the magnitude of bright fringe development is stronger at high temperature tetragonal phase (81$^{\circ}$C – 100$^{\circ}$C) compared to low temperature monoclinic phase (21$^{\circ}$C-59$^{\circ}$C). In addition, the change in Fresnel contrast from bright to dark as well as dark to bright at 63$^{\circ}$C and 77$^{\circ}$C indicates the start and end temperatures of monoclinic-tetragonal transition during heating . While the abrupt upward deflection around ~68$^{\circ}$C correlates with the studied structural phase transition of VO$_2$, the smooth downward deflections away from the phase boundary may arise due to thermal expansion mismatch between VO$_2$ and Si. Note that similar experiments performed on a control bare oxidized Si substrate over identical temperature range shows linear variation with no discontinuity in the defocus-temperature plot. The estimated stress-temperature plot (Fig.8d) shows an abrupt transformation stress of ~ 250 MPa (tensile) across phase transition on heating. This tensile stress development across the monoclinic-tetragonal phase transition is consistent with the *in situ* wafer curvature measurements carried out on the same VO$_2$ thin film that was deposited on a 4 inch wafer (Fig.2a). The higher values of slope observed for metallic phase (-3.08) compared to the insulting phase (-1.67) of stress-temperature plot (Fig.8d) may be suggestive of higher compressive stress development in metallic phase due to the larger thermal expansion mismatch with substrate.

## Conclusions

*In-situ* Fresnel contrast imaging can be utilized to visualize and quantify actuation induced by structural transitions. Abrupt, reversible cantilever tip deflection was found to be consistent with the estimated tensile stress relaxation across the monoclinic-tetragonal structural transition in $VO_2$. No time or drive rate dependence is observed in the stress relaxation indicating the athermal nature of phase transition kinetics. The study presented here can be widely applied to *in-situ* studies on solid-state phase transitions and could be of relevance to advancing the understanding of nano-electromechanical switches, environmentally responsive materials and actuation for robotic systems utilizing structural transitions driven by external stimuli.

## Experimental Section

### Thin film growth and cantilever fabrication

Thin films of $VO_2$ were grown on n-type Si (100) single crystal substrates by RF sputtering at $550^oC$ from a $V_2O_5$ target. Thin film stress measurements were carried out using TENCOR FLX-2320 instrument by *in-situ* wafer curvature from $50^oC$ to $110^oC$. Several heating-cooling thermal cycling was carried out with different heating rates varying from $2^oC/min$ to $30^oC/min$. Resistance was obtained from current-voltage (I- V) measurements using Keithley 236 Source Measure Unit in a temperature controlled probe station. To fabricate electron transparent $VO_2$/Si cantilevers, a two step method involving plan view sample preparation and FIB assisted milling is employed. First the TEM plan view sample is prepared by mechanical thinning down to 30 micron with diamond lapping films followed by Ar ion milling from substrate side until perforation. Finally, from the electron transparent region, $VO_2$ cantilevers are fabricated using FIB operated at 30kv and 80 pA current for milling. In situ TEM imaging and diffraction studies were carried out using JEM-2100 LaB6 Transmission Electron Microscope operated at 200 KV.

**Cantilever deflection measurement and Stress estimation**

The cantilever deflection is measured by careful observation of defocus values against temperature. In order to minimize errors caused by microscope instability in the cantilever deflection measurement, we optimized the conditions such as heating rate during the *in-situ* TEM studies. While the higher heating rate increases the temperature instability issues, lower heating rate increases the total time of the experiment making the precise cantilever deflection measurement difficult. Optimal heating rate of $2^oC$ to $5^oC$/min yielded consistent results. In all the cases the specimen is loaded such that the thin film side faces up during the imaging. The defocus values are calibrated by performing defocus imaging and "z" variation experiments with holey carbon grids and electron transparent bare Si specimens. From beam theory[30], the radius of curvature (R) is estimated from the tip deflection($\delta$) of the $VO_2$ cantilever of 5 μm length (L) using the following expression.

$$\delta = \frac{L^2}{2R} \quad\quad\quad\quad\quad\quad\quad (1)$$

With the knowledge of the mechanical properties and the physical thicknesses of cantilever components ($VO_2$ and Si), the stress change across the phase transition is estimated using the following modified Stoney's equation.[31]

$$\sigma = \frac{1}{6Rt_{VO_2}} \left[ \left(\frac{E_{VO_2}}{1-\upsilon_{VO_2}}\right) \frac{t^3_{VO_2}}{t_{VO_2}+t_{Si}} + \left(\frac{E_{Si}}{1-\upsilon_{Si}}\right) \frac{t^3_{Si}}{t_{VO_2}+t_{Si}} \right] \quad (2)$$

Where E, υ and t refers to elastic modulus, Poisson's ratio and thickness of Si and $VO_2$ layers noted as subscripts respectively. The elastic modulus of $VO_2$ is taken to be 140 GPa[32,33] with the assumed Poisson ratio of 0.3.[34] Radius of curvature (R) is calculated from the measured deflections of cantilever of known dimensions as described earlier. Similarly, the smooth cantilever deflection in the range of (25°C to 60°C) and (80°C-105°C) was further analyzed considering the thermal expansion mismatch stress of insulating and metallic $VO_2$ phases against Si substrate.

## Acknowledgements

Financial support from the National Science Foundation (Grant CCF-0926148) is acknowledged. This work was performed in part at the Center for Nanoscale Systems (CNS), Harvard University, a member of the National Nanotechnology Infrastructure Network (NNIN), which is supported by the National Science Foundation (NSF). We thank Dr. Changhyun Ko and Mr. Nicholas Antoniou for their assistance and technical support. We are much grateful to Prof. Frans Spaepen for valuable technical discussions.

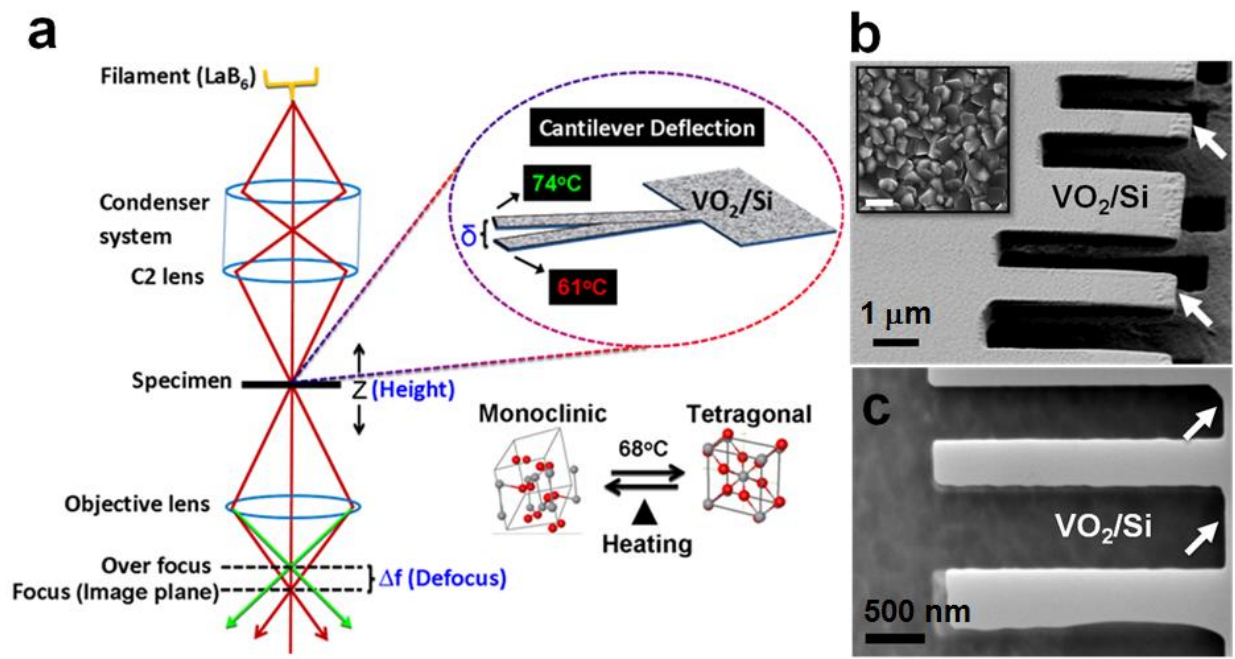

**Fig.1** Measuring phase transition induced cantilever deflection via in-situ TEM (a) Sketch of electron optics ray diagram[35] demonstrating *in-situ* Fresnel contrast imaging technique for temperature driven cantilever deflection measurement. Electron transparent $VO_2$/Si cantilevers are utilized to investigate the actuation process. Across the monoclinic-tetragonal transition, the cantilever shows upward deflection during heating and downward deflection on cooling. Inset shows the schematic of structural transition from monoclinic to tetragonal. The respective V and O atoms are marked as red and gray spheres respectively. Representative, low magnification SEM and TEM bright fright field images of $VO_2$/Si cantilevers are shown in Fig.1b and Fig.1c respectively. Zoomed SEM image (scale bar, 200 nm) of $VO_2$ cantilever shows the microstructure of $VO_2$ film (inset in Fig.1b).

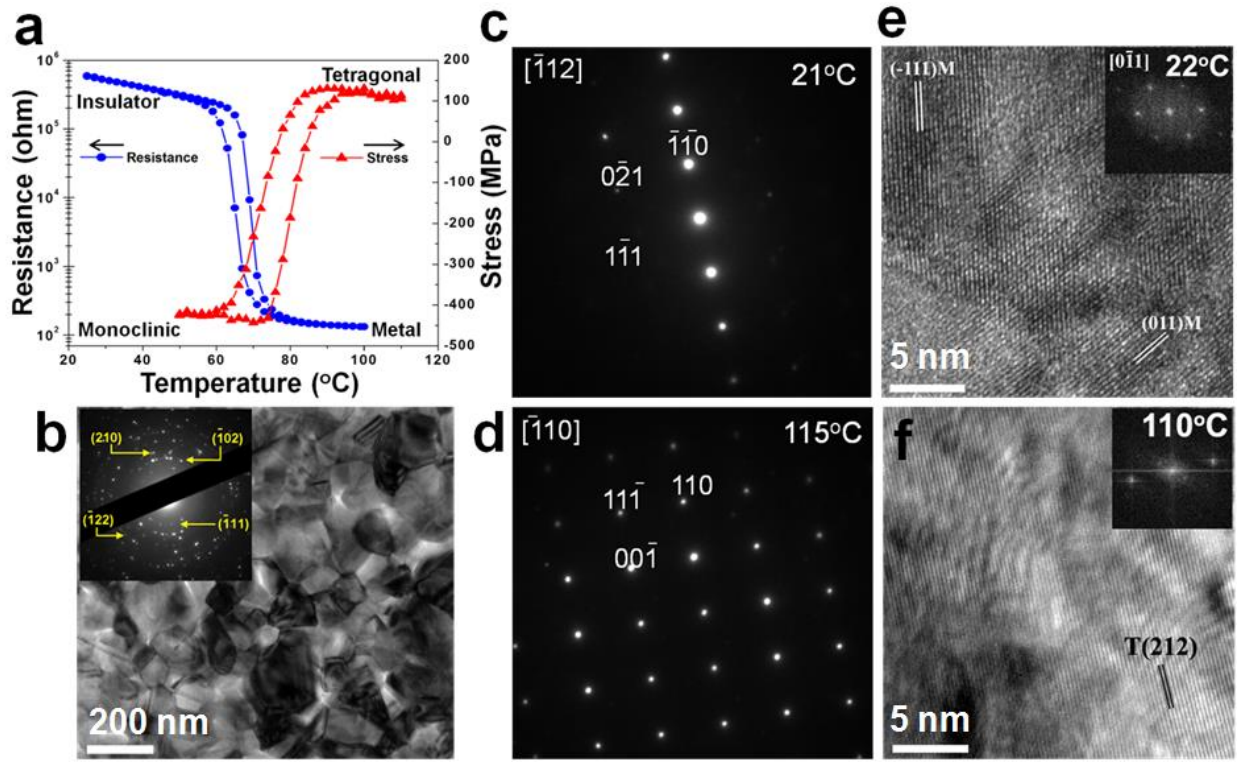

**Fig.2** Electrical, structural and stress relaxation characteristics of thermally driven metal insulator transition in VO$_2$ thin film. (a) Electrical resistance and in-situ wafer curvature stress measured across monoclinic-tetragonal phase transition during heating-cooling cycle. (b) TEM plan view bright field image of polycrystalline VO$_2$ film grown on Si (100) substrate. The corresponding selected area diffraction pattern is shown as inset. In situ electron diffraction pattern recorded from single grain showing (c) monoclinic [-112] and (d) tetragonal [-110] phases at 21$^o$C and 115$^o$C respectively. Lattice images recorded at high magnification showing monoclinic and tetragonal grains of VO$_2$ at (e) 22$^o$C and (f) 110$^o$C respectively consistent with the in situ electron diffraction studies across phase transition. The corresponding FFT patterns are shown as inset.

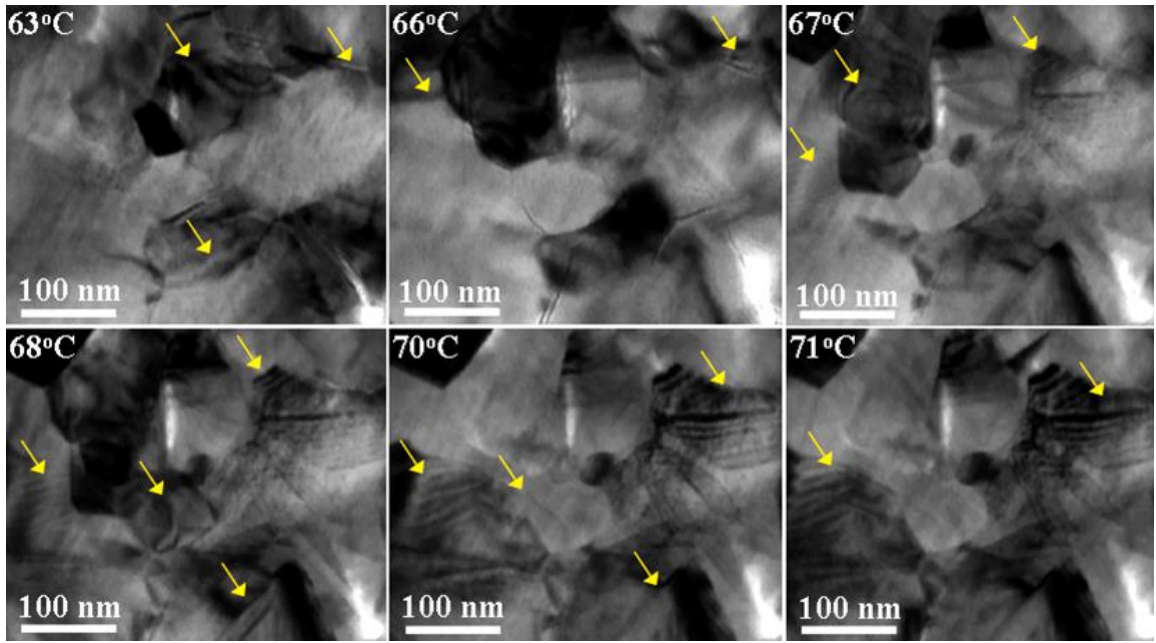

**Fig.3** *In-situ* TEM bright field observations across the phase transition boundary showing dynamic variation in step like contrast in VO$_2$ film. The development of dark fringes at the grain boundaries is evident at 66$^o$C.

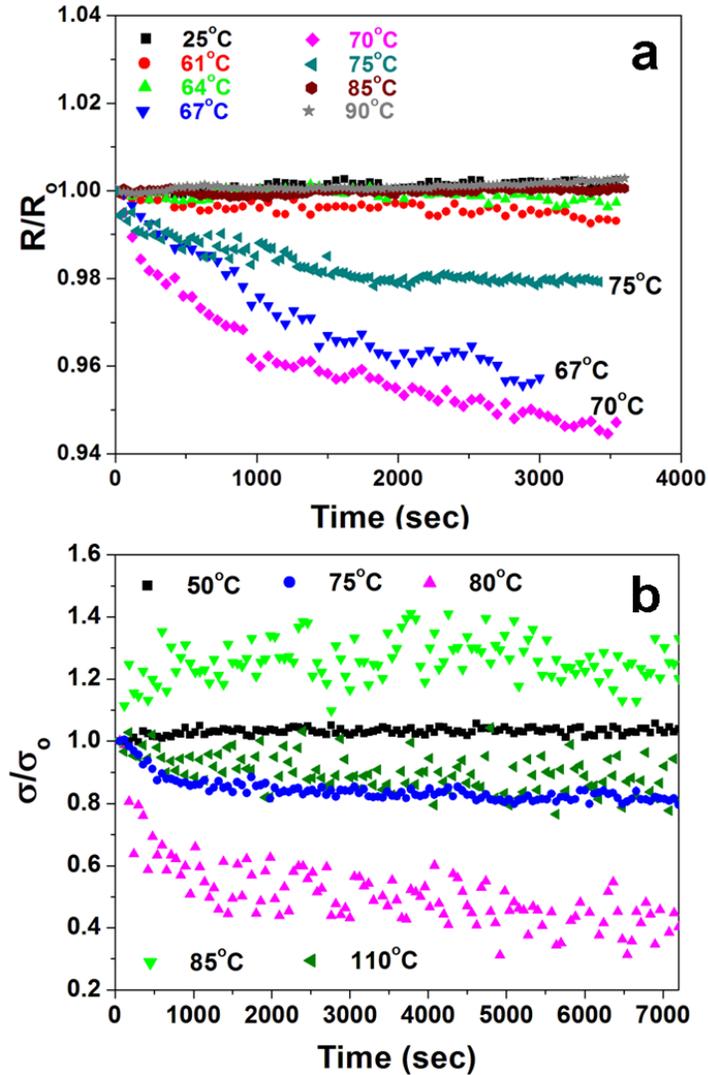

**Fig.4** Athermal phase transition characteristics in structural and electrical components (a) Resistance versus time plot showing the athermal phase transition of polycrystalline $VO_2$ thin films grown on Si (100) substrate. The overlapping plots of normalized resistance with initial time (time independent resistance) at different temperatures confirm the athermal nature of phase transition. Negligible fraction of time dependence observed at phase transition boundary of $67^{o}C$ and $70^{o}C$ is due to weak isothermal component of transition. Similar time independent phase transition is observed in stress relaxation measured from $VO_2$ thin film deposited on 4 inch Si (100) wafer as shown in (b)

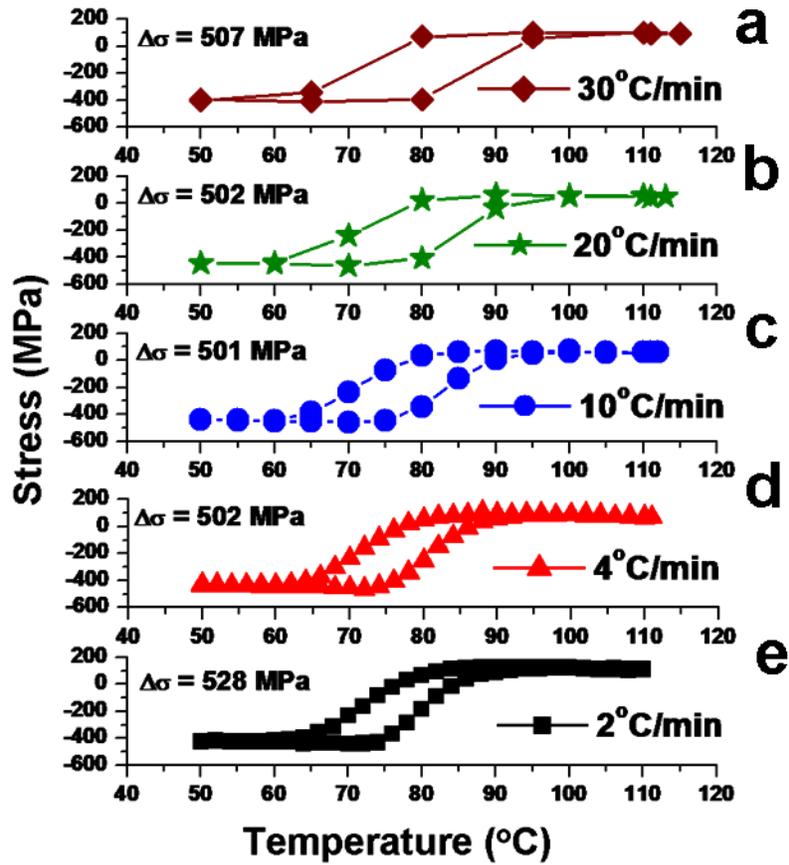

**Fig.5** Heating rate independent phase transition characteristics of $VO_2$ thin films. Stress relaxation across phase transition of $VO_2$ thin film at different heating rates measured by wafer curvature (a) 30°C/min, (b) 20°C/min, (c) 10°C/min, (d) 4°C/min and (e) 2°C/min showing the constant transformation stress of ~ 500 MPa.

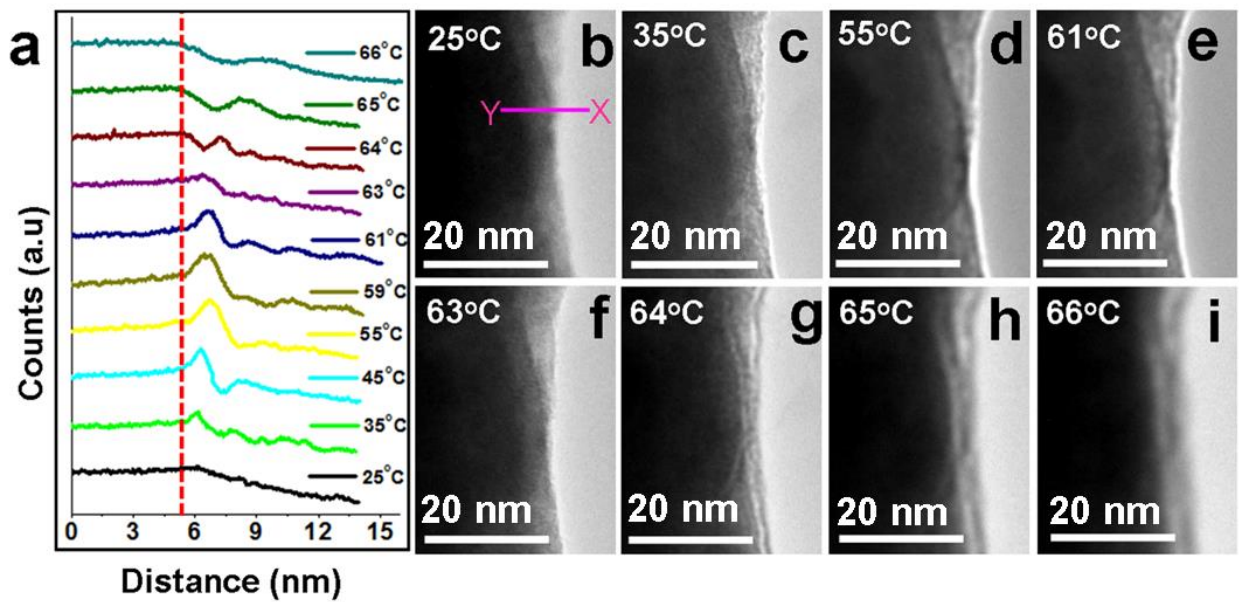

**Fig.6** In situ TEM images across the phase transition boundary showing variation in Fresnel contrast during heating. The observed bright fringe development (25°C to 61°C) and dark fringe development (above 64°C) indicates the cantilever deflection. Changes in the Fresnel contrast at the interface are more evident from the line profile analysis shown in Fig.6a. The corresponding images recorded at different temperatures from cantilever edge are shown in Fig.6b-i.

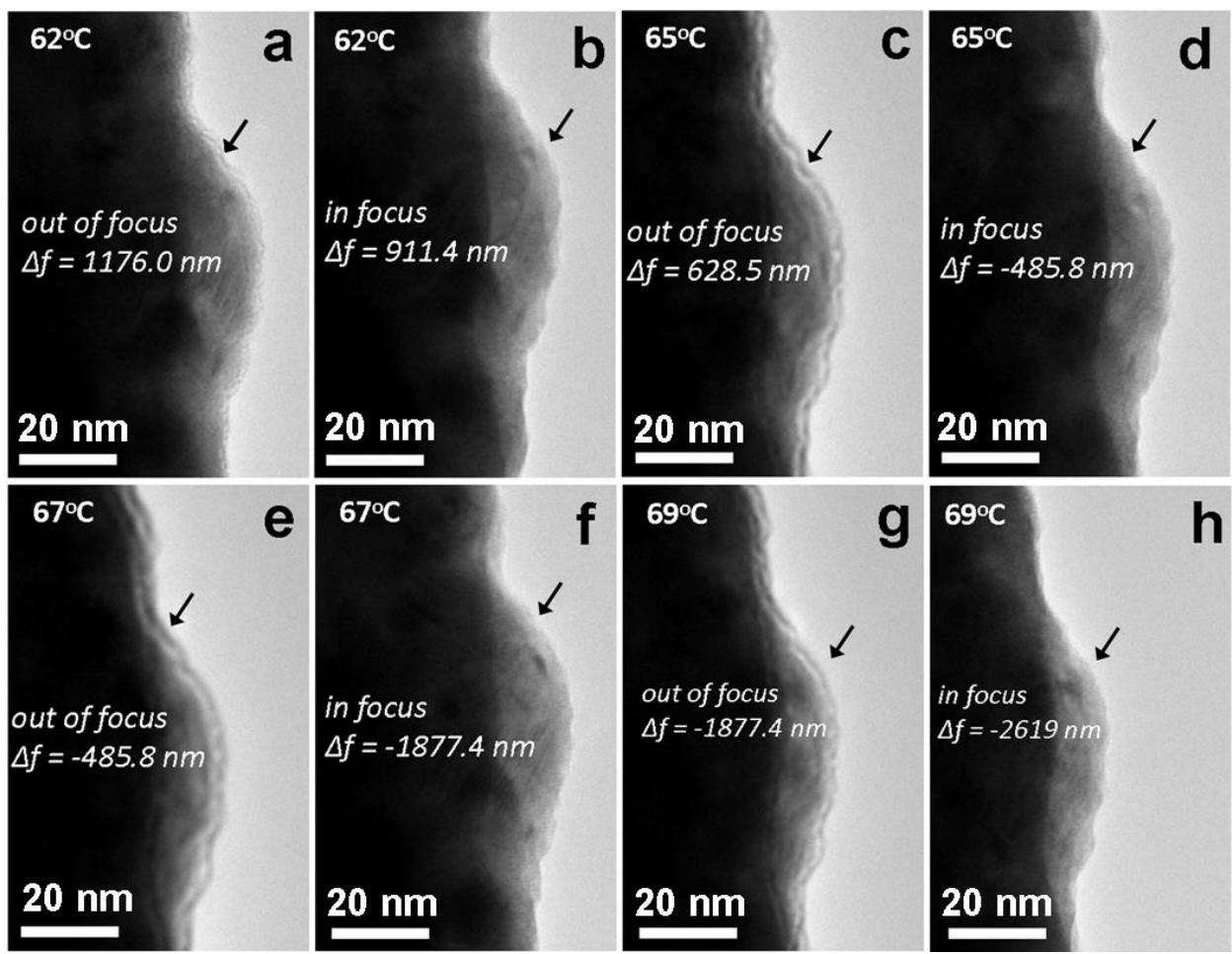

**Fig.7** In situ Fresnel contrast imaging across the phase transition. Dark fringe development across the phase transition during heating is used for cantilever deflection measurement. The corresponding defocus values used for iterative focusing are labeled. This captures the upward deflection of $VO_2$ cantilever in the temperature range of $62^oC$ to $74^oC$.

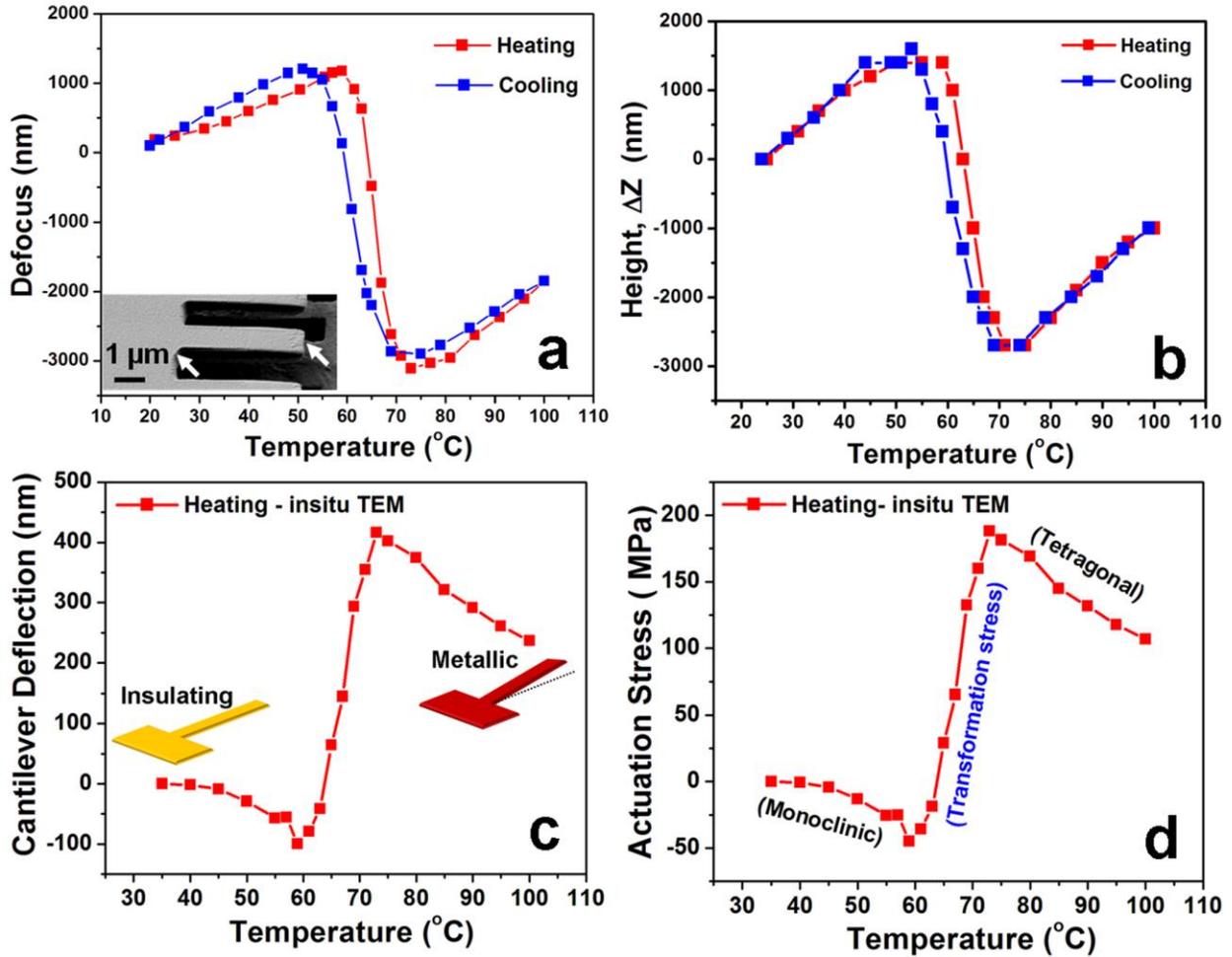

**Fig.8** Phase change-induced actuation of $VO_2$ cantilever captured by in situ Fresnel contrast imaging. (a) Actuation measured using defocus variation from the cantilever edge against temperature. Arrow marks in the SEM image of $VO_2$ cantilever show the cantilever edge and base used for in situ Fresnel contrast imaging (inset). (b) Specimen height variation against temperature shows the reversible actuation during heating-cooling cycle with hysteresis consistent with the defocus-temperature plot. (c) Actual cantilever tip deflection across the phase transition taking the deflection difference between the cantilever base and tip. (d) Stress development across the phase transition estimated from the measured cantilever deflection. Note that the estimated transformation stress from micron scale cantilever is consistent with the wafer scale actuation stress measurement shown in Fig.2a.